\documentclass[apj]{emulateapj}

\usepackage{amsmath,amstext,amssymb}
\usepackage{graphicx}
\usepackage{hyperref}
\usepackage[figure,figure*]{hypcap}
\usepackage{enumerate}
\usepackage{multirow}
\usepackage{amsfonts}
\usepackage{amssymb} 
\usepackage{bm}
\usepackage{natbib}
\usepackage{courier}
\usepackage{xcolor}

\newcommand{\del}[1]{}
\newcommand{\comm}[1]{}

\shorttitle{CHIPS1911+4455}
\shortauthors{Somboonpanyakul et al.}

\altaffiltext{\MIT}{Kavli Institute for Astrophysics and Space Research, Massachusetts Institute of Technology, 77 Massachusetts Avenue, Cambridge, MA 02139}
\altaffiltext{\UCin}{Department of Physics, University of Cincinnati, Cincinnati, OH 45221, USA}
\altaffiltext{\INAF}{INAF, Osservatorio di Astrofisica e Scienza dello Spazio, via P. Gobetti 93/3, 40129 Bologna, Italy}
\altaffiltext{\Princeton}{Department of Astrophysical Sciences, Princeton University, Princeton, NJ 08544, USA}
\altaffiltext{\Oslo}{Institute of Theoretical Astrophysics, University of Oslo, P.O. Box 1029, Blindern, NO-0315 Oslo, Norway}
\altaffiltext{\CfA}{Harvard-Smithsonian Center for Astrophysics, 60 Garden St., Cambridge MA 02138, USA}
\altaffiltext{\Michigan}{Physics \& Astronomy Department, Michigan State University, East Lansing, MI 48824-2320, USA}

\def\MIT{1}
\def\UCin{2}
\def\Michigan{3}
\def\INAF{4}
\def\Princeton{5}
\def\Oslo{6}
\def\CfA{7}

\begin{document}

\title{The Clusters Hiding in Plain Sight (\textit{CH\MakeLowercase{i}PS}) survey: \\CHIPS1911+4455, a Rapidly-Cooling Core in a Merging Cluster}

\author{
Taweewat Somboonpanyakul\altaffilmark{\MIT},
Michael McDonald\altaffilmark{\MIT}, 
Matthew Bayliss\altaffilmark{\UCin},
Mark Voit\altaffilmark{\Michigan},
Megan Donahue\altaffilmark{\Michigan},
Massimo Gaspari\altaffilmark{\INAF,\Princeton},
H\r{a}kon Dahle\altaffilmark{\Oslo}, 
Emil Rivera-Thorsen\altaffilmark{\Oslo}, and
Antony Stark\altaffilmark{\CfA}
}

\begin{abstract}
We present high-resolution optical images from the \emph{Hubble} Space Telescope, X-ray images from the \emph{Chandra} X-ray Observatory, and optical spectra from the Nordic Optical Telescope for a newly-discovered galaxy cluster, CHIPS1911+4455, at $z=0.485\pm0.005$. CHIPS1911+4455 was discovered in the Clusters Hiding in Plain Sight (\emph{CHiPS}) survey, which sought to discover galaxy clusters with extreme central galaxies that were misidentified as isolated X-ray point sources in the ROSAT All-Sky Survey. With new \emph{Chandra} X-ray observations, we find the core ($r=10$\,kpc) entropy to be $17^{+2}_{-9}$ $\rm{keV~cm^2}$, suggesting a strong cool core, which are typically found at the centers of relaxed clusters. However, the large-scale morphology of CHIPS1911+4455 is highly asymmetric, pointing to a more dynamically active and turbulent cluster. Furthermore, the \emph{Hubble} images reveal a massive, filamentary starburst near the brightest cluster galaxy (BCG). We measure the star formation rate for the BCG to be 140--190 $\rm{M_{\odot}\,yr^{-1}}$, which is one of the highest rates measured in a central cluster galaxy to date. One possible scenario for CHIPS1911+4455 is that the cool core was displaced during a major merger and rapidly cooled, with cool, star-forming gas raining back toward the core. This unique system is an excellent case study for high-redshift clusters, where such phenomena are proving to be more common. Further studies of such systems will drastically improve our understanding of the relation between cluster mergers and cooling, and how these fit in the bigger picture of active galactic nuclei (AGN) feedback. 
\end{abstract}

\keywords{galaxies: clusters: general --- galaxies: clusters: intracluster medium --- X-rays: galaxies: clusters}

\section{Introduction}
Early X-ray observations of the intracluster medium (ICM) in the center of galaxy clusters revealed cooling times much shorter than the Hubble time, leading to the development of the cooling flow model \citep[e.g.,][]{Fabian1994}. In this model, hot gas in dense cores should radiatively cool and fuel 100--1000 $\rm{M_{\odot}}$/yr starbursts in the central brightest cluster galaxy (BCG). However, many studies have shown that BCGs are only forming stars at $\sim$1\% of this rate \citep[e.g.,][]{McDonald2018}. A promising mechanism proposed for preventing cooling of the ICM is AGN heating by jets and bubble-induced weak shocks~\citep[see reviews by][]{Fabian2012, McNamara2012,Gaspari2020}. Evidence supporting these theories includes the ubiquitous presence of radio galaxies at the center of clusters~\citep{Sun2009} and the similarity between the mechanical energy released by AGN driven bubbles and the energy needed to quench cooling~\citep[e.g.][]{Rafferty2008,Birzan2008,Hlavacek-Larrondo2015}.

Galaxy clusters with signatures of cooling in their centers are often called ``cool-core" (CC) clusters, with their counterparts being referred to as ``non cool-core" (NCC) clusters. \citet{Hudson2010} found that the best way to segregate the two is to consider their central cooling time ($\rm{t_{cool}}$). Specifically, CC clusters have $\rm{t_{cool}}<7.7$ Gyr, while clusters with $\rm{t_{cool}}<1.0$ Gyr are referred to as ``strong CCs''. A number of observational studies have found that CCs are mostly found in relaxed clusters while NCCs reside in dynamically-active clusters. Indeed, all of the strongest CCs known (based on cooling rate) are found in the most relaxed clusters (e.g., Phoenix~\citep{McDonald2012}, Abell 1835~\citep{McNamara2006}, Zw3146~\citep{Edge1994}). This is also consistent with a variety of other studies that found star-forming BCGs in the most relaxed CC clusters \citep{Crawford1999, Donahue2010, Molendi2016, Cerulo2019}. On the other hand, morphologically-disturbed clusters (which are likely to be recent mergers) generally have no evidence for ongoing cooling, suggesting that major mergers may have the potential to destroy cool cores \citep{Burns2008,Poole2008} through shock-heating \citep{Burns1997} and mixing \citep{Gomez2002}. 
 
The discovery of CHIPS1911+4455 (Somboonpanyakul et al., in press) runs counter to these established norms, since it not only harbors a very blue (star-forming) galaxy in the center, but also shows a highly-disturbed morphology on both large ($\sim$200 kpc) and small ($\sim$20 kpc) scales. There are no known nearby clusters that have properties similar to CHIP1911+4455, though \citet{McDonald2016} reports a higher fraction of star-forming BCGs in merging clusters at $z>1$. This implies that CHIPS1911+4455 may provide an avenue for studying a high-redshift phenomenon in a low-redshift cluster. To fully understand this system we have obtained new observations in the core of CHIPS1911+4455, which we will discuss below.

Throughout this paper, we assume $\rm{H_0}=70\,\rm{km\,s^{-1} Mpc^{-1}}$, $\Omega_{m}=0.3$, and $\Omega_{\Lambda}=0.7$. All errors are $1\,\sigma$ unless noted otherwise.

\section{Observations}~\label{sec::data}
In this section, we summarize the acquisition and reduction of data obtained from the \emph{Chandra} X-ray telescope, the \emph{Hubble} Space Telescope and the Nordic Optical Telescope (NOT). 

\subsection{X-ray: Chandra}
\begin{figure*}
	\begin{center}
		\begin{tabular}{cc}
			\includegraphics[width=1\linewidth]{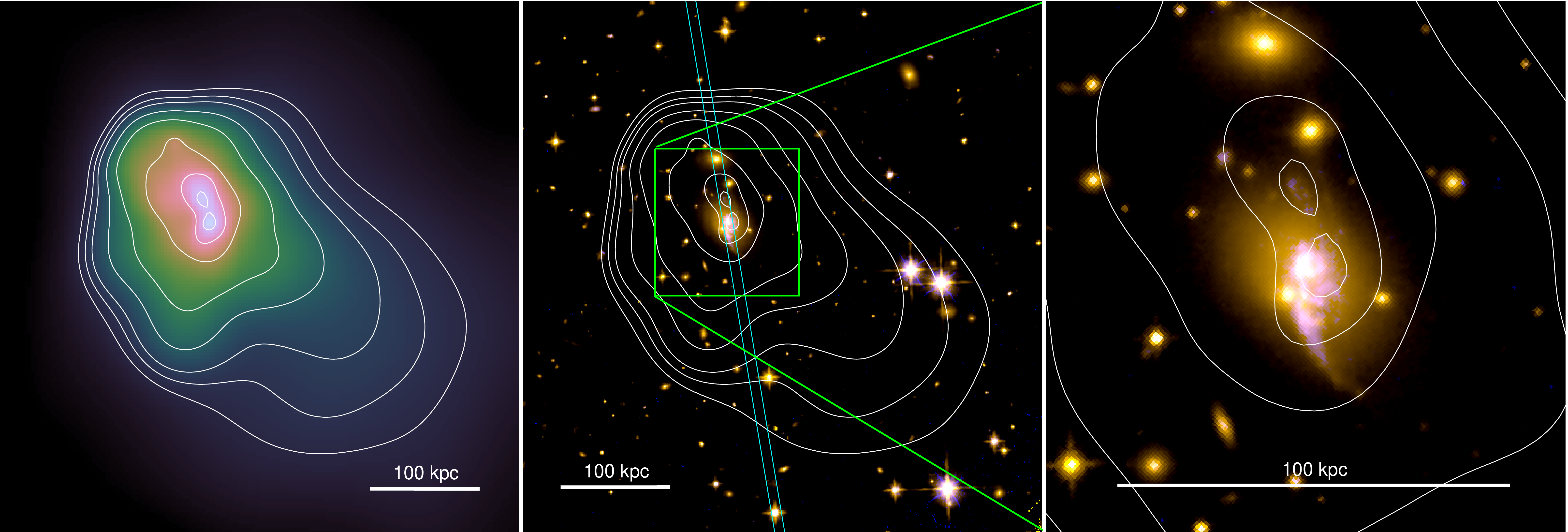}
		\end{tabular}
		\vspace{-0.05in}
		\caption{Left: \emph{Chandra} 0.5--7.0 keV image of CHIPS1911+4455, highlighting the asymmetric morphology on both small and large scales. The image is oriented such that North is up and East is to the left. Middle: \emph{Hubble} images with X-ray contours overlaid. The contour lines were chosen arbitrarily to guide the eye. The cyan box shows the orientation of the long slit (see also Fig.~\ref{fig::spectrum}). Right: The \emph{Hubble} images of the central galaxy, showing the blue star-forming filaments, extending on scales of $\sim30$ kpc. These images show that the cool, star-forming gas is centered on the X-ray peak, with a faint set of filaments extending north to the secondary peak and a brighter filament extending to the south.}
		\label{fig::chandra_hst}
	\end{center}
\end{figure*}

CHIPS1911+4455 (OBSID: 21544) was observed in 2019 with \emph{Chandra} ACIS-I for a total of 30.5 ks. The data were analyzed with \verb|CIAO| v4.11 and \verb|CALDB| v4.8.5 and recalibrated with VFAINT mode for improved background screening. To look for small-scale structures near the center of the cluster, images were smoothed adaptively, using \verb|CSMOOTH|\footnote{https://cxc.harvard.edu/ciao/ahelp/csmooth.html} to achieve a uniform signal-to-noise ratio over the full image, as shown in the left panel of Fig.~\ref{fig::chandra_hst}.

The temperature profile was extracted from coarse annuli so that the number of counts per annulus was around 800, which is enough to get well-constrained temperature measurements ($\Delta \rm{kT}/\rm{kT}$ $\sim$20\%). All spectra were fit simultaneously with the \verb|APEC| model for the cluster emission, a second \verb|APEC| model for the Milky Way ($kT=0.18$\,keV), the \verb|PHABS| model for Galactic absorption, and the \verb|BREMSS| model to represent a hard background ($kT=0.40$\,keV) from unresolved point sources, following~\cite{McDonald2013}. The \textit{WSTAT} statistic was used.

The gas density profile was created by first computing the 0.7-2.0 keV surface brightness profile. The conversion from the X-ray surface brightness profile to the emission-measure profile (EM($r$) = $\int n_{p}n_{e}dl$) was calculated as a function of radius based on the best-fit temperature profile and assuming a collisionally-ionized plasma \verb|APEC| model with metallicity 0.3Z$_{\odot}$. For more details of the X-ray analysis, see \cite{Somboonpanyakul2018}.

\subsection{Optical: Hubble}
CHIPS1911+4455 was observed for two orbits with the \emph{Hubble} Space Telescope (HST) during Cycle 27. The data include medium band F550M data from the Advance Camera for Surveys (ACS) and broad band F110W data from the Wide Field Camera infrared channel (WFC3-IR). The F550M filter contains both the blue continuum and the bright [O II] doublet at the redshift of the cluster, which should both be elevated in star-forming regions. The F110W filter, on the other hand, is sensitive to the red continuum, probing the old stellar populations of the BCG and other cluster members.

\subsection{Optical Spectra: Nordic Telescope}
Two optical spectra of the BCG of CHIPS1911+4455 were obtained with the Alhambra Faint Object Spectrograph and Camera (ALFOSC) at the 2.56m Nordic Optical Telescope (NOT) on May 9th, 2019. One of the spectra was obtained from Grism$\#4$ (R=360, 3200--9600${\rm \AA}$) with 1.3$^{\prime\prime}$ slit for 1500-second exposure. The other spectrum was a stack of two 1100-second spectra from Grism$\#5$ (R=415, 5000--10700${\rm \AA}$) with 1.3$^{\prime\prime}$ slit at $90^\circ$ from the first spectrum. Wavelength solutions for the two spectra were calibrated with HeNe and ThAr arc lamps, respectively, with an absolute calibration uncertainty of 2\AA. Masks were applied to remove cosmic-rays before the 1D spectra were extracted from the 2D spectral images. The 1D spectra were then flat-fielded, and background-subtracted from off-source regions surrounding the 1D extraction region.

\section{Results}~\label{sec::result}
\subsection{CHIPS1911+4455: A Strong Cool Core}\label{sec::coolcore}
\begin{figure*}
	\begin{center}
		\includegraphics[width=0.65\columnwidth]{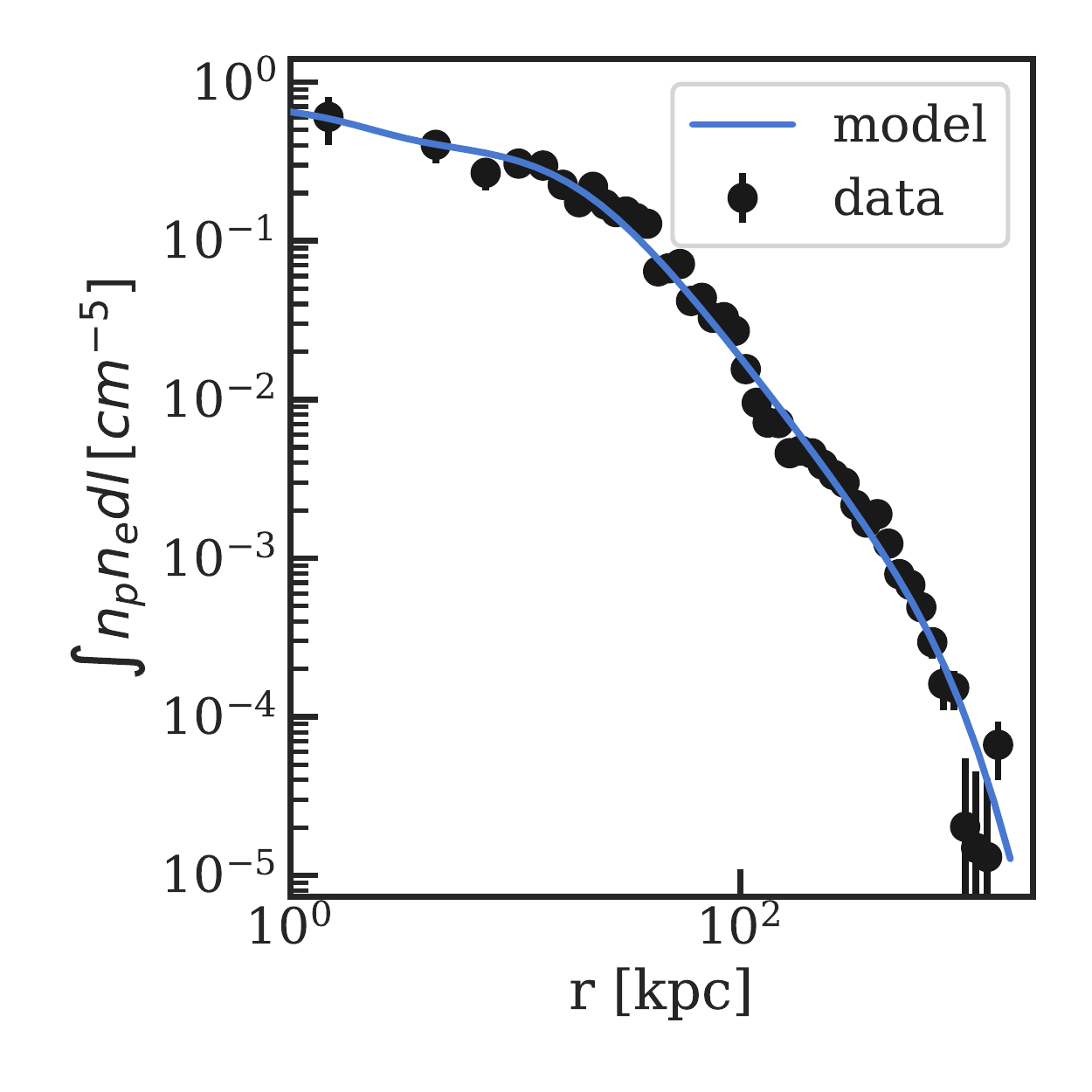}
		\includegraphics[width=0.65\columnwidth]{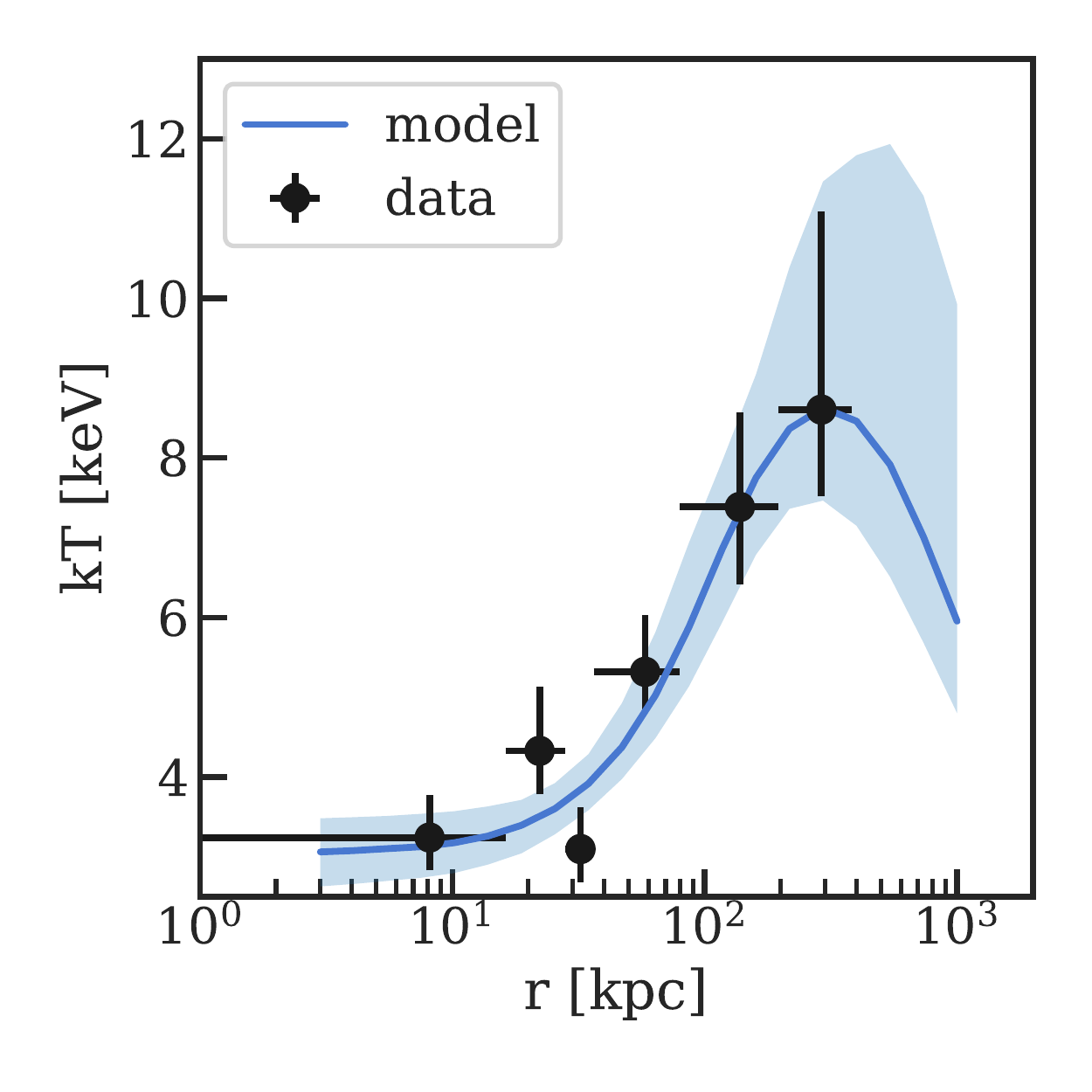}
		\includegraphics[width=0.65\columnwidth]{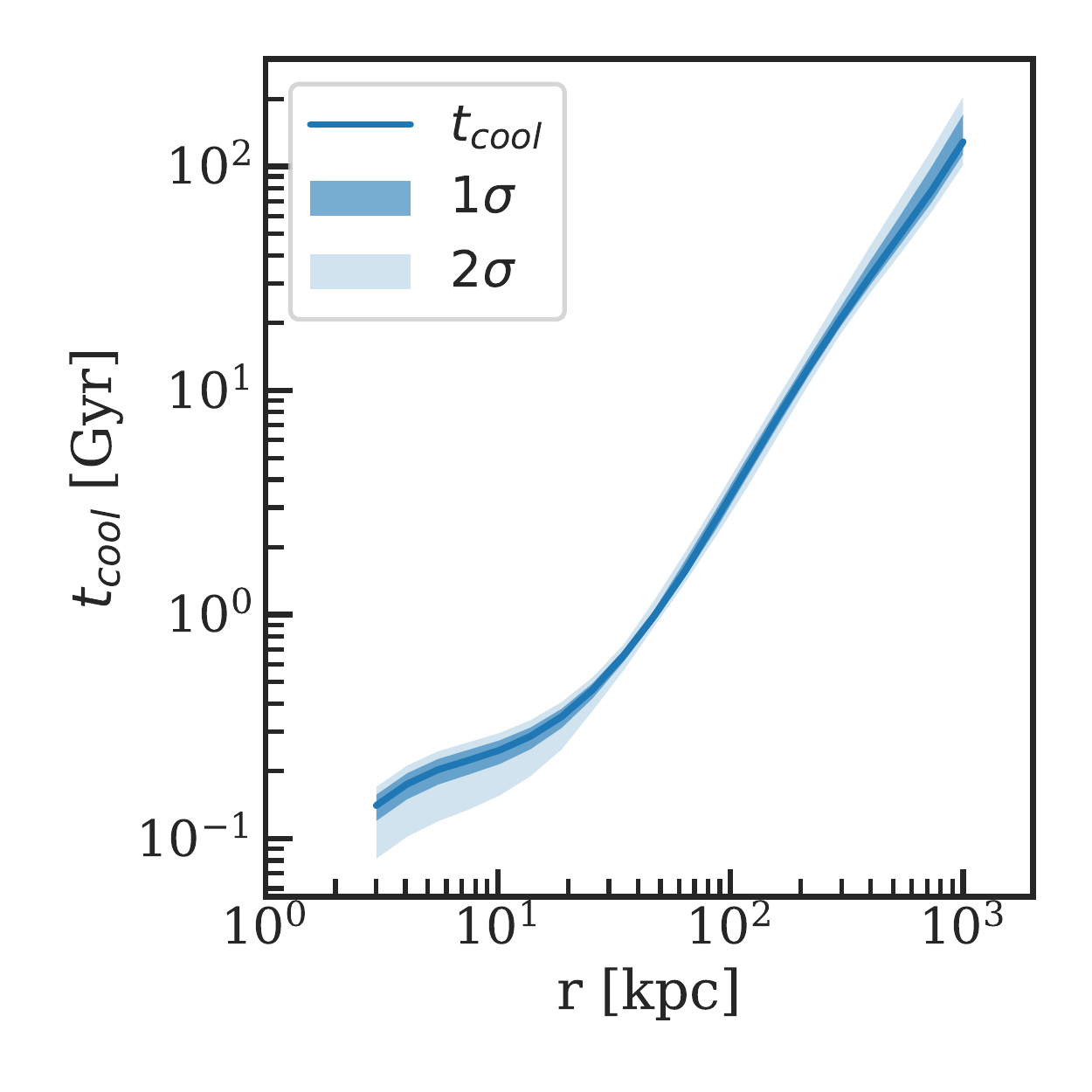}
		\caption{Left: The surface brightness profile of CHIPS1911+4455. The black dots are data points, and the blue line is the best-fit model \citep[see][for a further description of this modeling]{Somboonpanyakul2018}. Middle: the projected 2-D temperature profile of the cluster. The black dots are data extracted from modeling of the X-ray spectra. The blue line is the best-fit model following \cite{Vikhlinin2006}. Right: The cooling time profile of the cluster. The cyan-shaded region corresponds to $1\sigma$ credible region from the models in the left and center panels, while the light blue-shaded region corresponds to $2\sigma$ credible region. This cluster is classified as a strong cool core, based on a large drop in the core temperature and a central cooling time less than 1 Gyr~\citep{Hudson2010}.}
		\label{fig::temp_den}
	\end{center}
\end{figure*}

The thermodynamic profiles of CHIPS1911+4455 are shown in Fig. ~\ref{fig::temp_den}. The electron density at 10 kpc is $0.0884\,\rm{cm^{-3}}$, which is among the highest measured to date~\citep{Hudson2010,McDonald2017}, while the temperature profile drops from a maximum of $\sim$8 keV at $\sim$300 kpc to 4 keV at $\sim$10 kpc.

\begin{figure}
	\begin{center}
		\includegraphics[width=0.95\columnwidth]{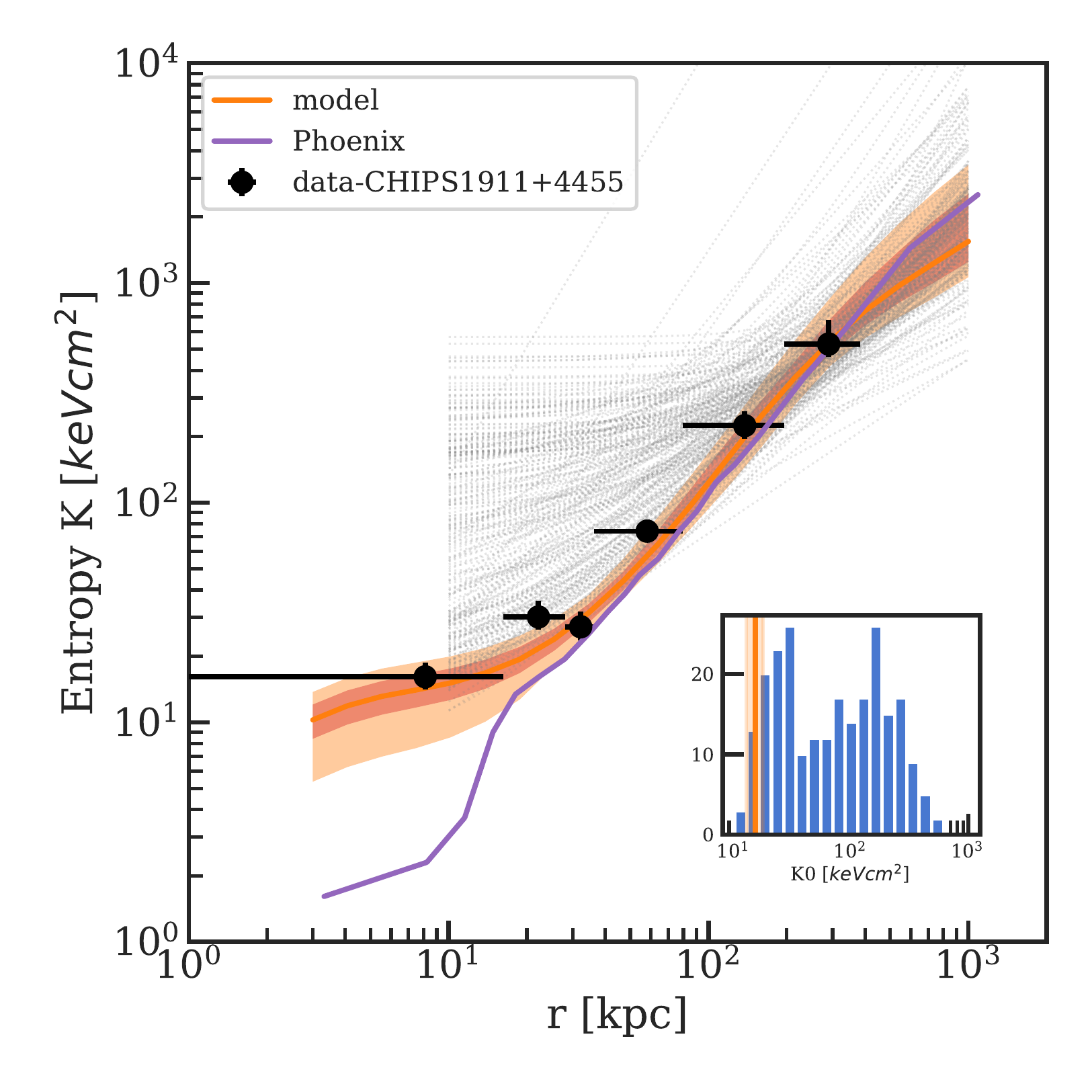}
		\caption{Entropy profile for CHIPS1911+4455 (orange), compared to 239 clusters from the ACCEPT survey~\citep[gray;][]{Cavagnolo2009} and the Phoenix cluster~\citep[purple;][]{McDonald2019}. The shaded region corresponds to $1\sigma$ and $2\sigma$ credible region (see Fig.~\ref{fig::temp_den}). The black dots are data points, estimated from the projected 2D temperature data and the 3D density model. The inset shows the histogram of the core entropy ($r < 10$\,kpc) of all the ACCEPT clusters and CHIPS1911+4455 (orange). CHIPS1911+4455's core entropy is in the lowest 10\% of all ACCEPT clusters.}
		\label{fig::entropy}
	\end{center}
\end{figure}

The entropy of the ICM ($K \equiv kT n_e^{-2/3}$) reflects the thermal history of a cluster, which is solely affected by heat gains and losses~\citep{Cavagnolo2009,Panagoulia2014}, while the cooling time ($t_{cool}\equiv\frac{3}{2}\frac{(n_e+n_p)kT}{n_en_p\Lambda(T)}$) represents the amount of time required for the ICM to radiate all the excess heat, where $\Lambda(T)$ is the cooling function~\citep{Sutherland1993}. For CHIPS1911+4455, the central ($r=10$\,kpc) cooling time is $98^{+7}_{-32}\,\rm{Myr}$, which is classified as a strong cool core~\citep{Hudson2010}. The deprojected entropy profiles for CHIPS1911+4455, the Phoenix cluster, and hundreds of clusters from the ACCEPT survey~\citep{Cavagnolo2009} are shown in Fig.~\ref{fig::entropy}. Both CHIPS1911+4455 and the Phoenix cluster have entropy profiles that are among the lowest known. The core ($r=10$\,kpc) entropy for CHIPS1911+4455 is $17^{+2}_{-9}\,\rm{keV cm^2}$, which is in the 5$^{th}$ percentile of all clusters in the ACCEPT survey.

\begin{figure}
	\begin{center}
		\includegraphics[width=0.95\columnwidth]{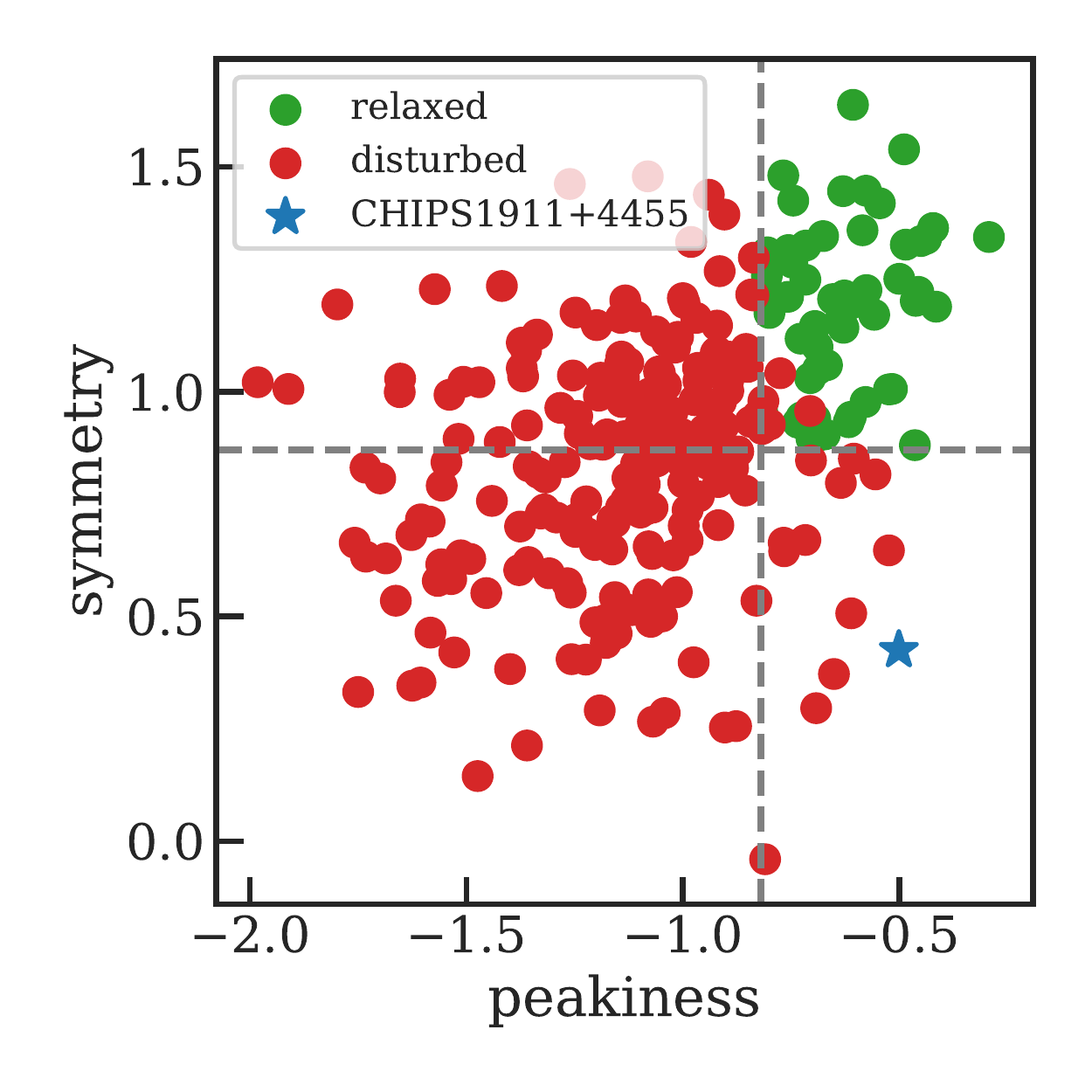}
		\caption{Distribution of X-ray symmetry and peakiness for clusters presented in~\citet{Mantz2015a}. Dashed lines show the cuts used to define the relaxed sample, which were chosen to broadly agree with ``by-eye'' assessments of relaxedness. Clusters satisfying this criterion are shown in green, while red points show the non-relaxed clusters. The blue star represents the location of CHIPS1911+4455, which is the peakiest cluster with large-scale asymmetry.}
		\label{fig::symmetry}
	\end{center}
\end{figure}

\subsection{CHIPS1911+4455: A Major Merger}\label{sec::morp}
There are various ways to quantify the dynamical state of a cluster, with some of the best combining X-ray data with information on the galaxy velocities. Lacking the latter, we restrict ourselves to X-ray-only indicators. The two particular quantities we consider are the ``peakiness'' of the surface brightness profile and the distance between the center of symmetry on small and large scale (``symmetry''), following \citet{Mantz2015a}. The peakiness measure is a proxy for the presence of a cool core, which are typically found in relaxed clusters. A cluster with high symmetry appears similar on small and large scales, suggesting that it is dynamically relaxed. Given that both of these proxies probe the dynamical state of the cluster, albeit in different ways, it is unsurprising that they are correlated~\citep{Mantz2015a}. The green points in Fig.~\ref{fig::symmetry} show the population of relaxed clusters in this morphology plane. For CHIPS1911+4455, the peakiness is measured to be -0.501, which is in the $96^{\rm{th}}$ percentile of all clusters, while the symmetry is estimated to be 0.425, which is in the $7^{\rm{th}}$ percentile. The fact that CHIPS1911+4455 is simultaneously one of the strongest cool cores \emph{and} least symmetric clusters known is highly unusual.

\subsection{CHIPS1911+4455: A Starburst BCG}\label{sec::SFR}
In Fig.~\ref{fig::chandra_hst}, we compare optical and X-ray images of CHIPS1911+4455. The \emph{Hubble} images show that the red emission from old stellar populations is relatively smooth and symmetric, while the blue emission from the young stellar populations and cool gas is clumpy, asymmetric, and extended on $>$30 kpc scales. In particular, the red emission shows no elongation towards the galaxies to the north, suggesting that there has not been a recent interaction between these galaxies. Given this, and the fact that the central entropy is amongst the $\sim$2--3 lowest ever measured, we propose that the young stars are forming directly from the cooling ICM. The filamentary complex structures in the blue emission are similar to the emission nebula in the Phoenix cluster~\citep{McDonald2019} and other nearby cool core clusters~\citep[e.g.,][]{McDonald2011,Tremblay2015,Olivares2019}.

\begin{figure*}
	\begin{center}
		\includegraphics[width=1.97\columnwidth]{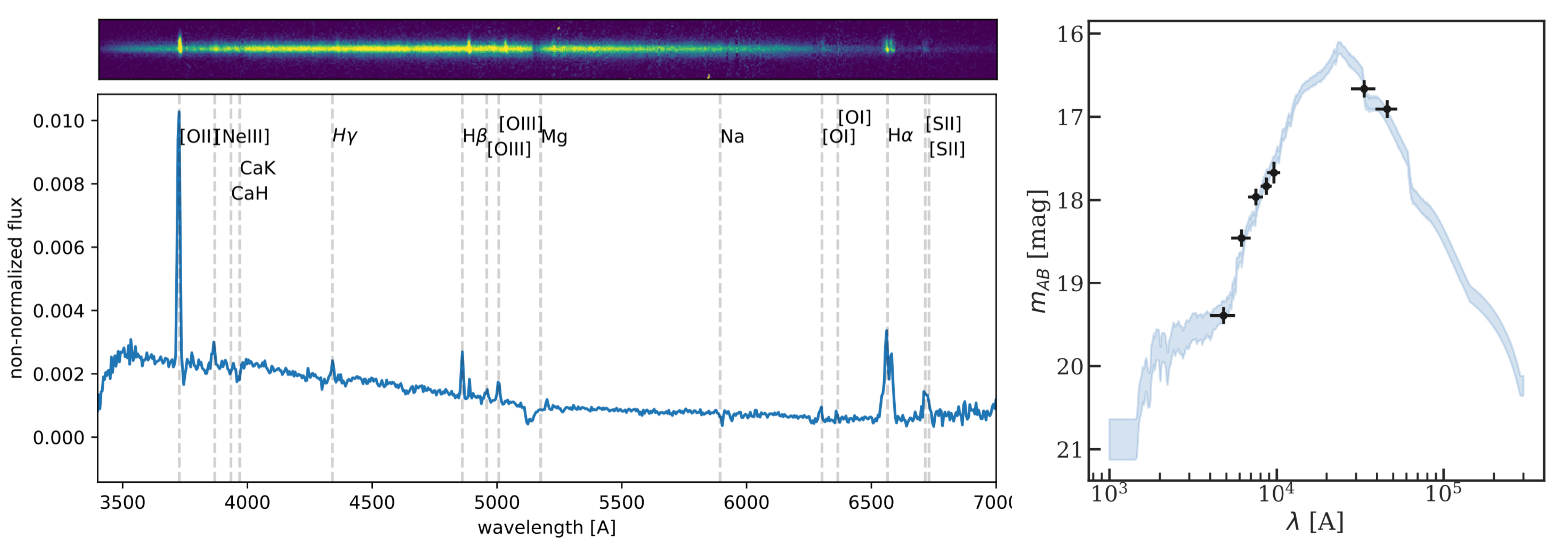}
		\caption{Top left: The 2D spectral image of the BCG of CHIPS1911+4455 from the Nordic Optical Telescope showing that the line-emitting gas is extended along the slit (vertical) direction. This confirms that the [O II] emission is not coming (exclusively) from a central AGN, and that the extended emission in Fig.~\ref{fig::chandra_hst} is indicating the presence of cool gas, rather than purely blue continuum from young stars.
		Bottom left: The 1D spectrum in the rest frame from the 2D spectral image above. Grey dash lines show the location of well-known emission lines. The spectrum clearly shows the strong [O II] doublet at 3727$\rm{\AA}$ and the $\rm{H{\alpha}}$ emission line at 6562.8$\rm{\AA}$. Right: The fitted spectral energy distribution (SED) model with broad-band optical data ($g$, $r$, $i$, $z$, and $y$) from Pan-STARRS and mid-IR from WISE.}
		\label{fig::spectrum}
	\end{center}
\end{figure*}

From the optical spectrum in Fig.~\ref{fig::spectrum}, we identify several bright emission lines, including H$\beta$ and [O\,\textsc{ii}]. The relative lack of bright [O\,\textsc{iii}] compared to [O\,\textsc{ii}] indicates that the central galaxy in CHIPS1911+4455 is a massive starburst and not a bright AGN. We use these emission lines to estimate the redshift of the central galaxy, finding $z=0.485\pm0.005$. 
From the two slit orientations, we measure [O\,\textsc{ii}] equivalent widths of $40.9\pm1.0$ $\rm{\AA}$ and $43.8\pm1.0$ $\rm{\AA}$, which are consistent with one another. To convert to flux, we model the spectral energy distribution (SED) of the galaxy, based on data from Pan-STARRS ($g$, $i$, $r$, $z$, and $y$)~\citep{Tonry2012} and WISE ($w1$ and $w2$)~\citep{Wright2010}, with a linear combination of ``young" and ``old" stellar populations, along with dust reddening~\citep{Calzetti2000}. The best-fit SED model is shown in Fig.~\ref{fig::spectrum}. We estimate a de-reddened continuum flux at the location of [O\,\textsc{ii}] of $3.5\pm0.5\times10^{-16}\,{\rm erg\,s^{-1}\,cm^{-2}\,\AA^{-1}}$, which we combine with the EW to estimate the star formation rate (SFR) of the central galaxy, arriving at $189^{+25}_{-22}\,\rm{M_{\odot}\,yr^{-1}}$~\citep{Kennicutt1998}.

Another way to measure SFR is to use the 24$\mu m$ emission since mid-IR fluxes are unaffected by dust extinction, unlike UV and optical tracers. Instead, mid-IR emission comes from the reprocessed light by dust, produced from recently formed stars. Based on the WISE4 flux ($\sim 4\times10^{-18}\,{\rm erg\,s^{-1}\,cm^{-2}\,\AA^{-1}}$), we estimate the SFR for the central galaxy to be $143^{+31}_{-26},\rm{M_{\odot}\,yr^{-1}}$, based on \citet{Cluver2017}.

Considering both the [O\,\textsc{ii}] emission line luminosity and the mid-IR continuum, we obtain consistent SFRs of $\sim$140--190 $\rm{M_{\odot}\,yr^{-1}}$, making the BCG in CHIPS1911+4455 one of the most star-forming BCGs in the $z<1$ Universe~\citep[see e.g.,][]{McDonald2018}

\section{Discussion}~\label{sec::discussion}
It is clear, based on Sections~\ref{sec::coolcore}--\ref{sec::SFR}, that CHIPS1911+4455 is a unique system in the nearby universe ($z<0.5$) with an extremely high central SFR, a strong cool core, and large-scale morphology consistent with a recent merger. It is common knowledge that, at low-redshift, the most star-forming BCGs tend to be located in the most relaxed, cool-core clusters~\citep{Cavagnolo2008, Donahue2010}. Whereas, the SFR in high-redshift BCGs tend to be higher in dynamically active, non cool-core clusters~\citep{McDonald2016, Bonaventura2017}. CHIPS1911+4455 may represent a low-redshift analog of these high-redshift systems, allowing us to study physical processes that may have been much more common at early times. Below, we speculate on possible explanations for the observed starburst.

In the precipitation scenario \citep{Voit2015} for AGN feeding and feedback, cool clouds condense and can form stars or fuel AGN feedback when the cooling time ($t_{\rm{cool}}$) is significantly shorter than the timescale for the cloud to fall to the center of the cluster potential ($t_{\rm{ff}}$). The latter timescale is a proxy for the mixing time of the cooling gas. The threshold for thermal instability is usually taken to be $t_{\rm{cool}}/t_{\rm{ff}}<10$ \citep[e.g.,][]{Voit2015,Gaspari2020}, but this depends on the medium's susceptibility to condensation, the slope of the entropy profile, the amount of turbulence, and the amplitude of entropy perturbation~\citep{Voit2017,Voit2018}. 
If a dense, low-entropy core is perturbed from the center of the gravitational potential, the timescale for mixing is increased while the cooling time remains constant, leading to an enhancement in thermal instabilities. Further, the separation of the low-entropy gas from the direct influence of the central AGN could prevent t$_{cool}$ from increasing. This displacement of low-entropy gas from the central massive galaxy would happen naturally during a merger (i.e., in CHIPS1911+4455). The X-ray data in Fig.~\ref{fig::chandra_hst} support this scenario, depicting a disturbed cool core elongated in the north-south direction. While the bulk of the cool, star-forming gas extends along a southern-pointing filament, there is a fainter blue filament that connects the central BCG to the more northern X-ray peak. This suggests that the northern X-ray peak might contain low-entropy gas that is dislodged from the location of the BCG.
The existence of this system -- a dynamically active but rapidly cooling core -- provides evidence that mergers may stimulate star formation and enhance cooling. This process would be especially relevant in the distant universe when mergers were more common compared to present time~\citep{Fakhouri2010}.

Closely related to this, in the chaotic cold accretion scenario~\citep[CCA; e.g.,][]{Gaspari2019} turbulence is a key ingredient to drive direct non-linear thermal instability and extended condensation. During CCA the global driver can be tied not only to AGN feedback, but to mergers too, which can stimulate significant amount of turbulence~\citep[e.g.,][]{Lau2017} and density fluctuations at large injection scales, enabling enhanced condensation.

In summary, there are multiple ways that a recent cluster-cluster merger could enhance cooling: by increasing the fall-back time of a low-entropy cloud (lowering $t_{\rm{cool}}/t_{\rm{ff}}$), by physically separating the cool gas and the heat source (AGN), and by increasing the local turbulence, leading to larger density fluctuations. The inclusion of dynamically-active clusters in the suite of high-resolution, zoom-in simulations that study the detailed interplay between radio-loud AGN and the cooling, multiphase ICM \citep[e.g.,][]{Gaspari2019}, is a necessary next step towards understand the relative importance of these phenomena.

\section{Conclusion}~\label{sec::conclusion}
In this work, we present new data from \emph{Hubble}, \emph{Chandra}, and the Nordic Optical Telescope. Our findings are summarized as follows:

\begin{enumerate}
	\item We measure the ICM density in the core ($r<10$\,kpc) of CHIPS1911+4455 to be 0.09 $\rm{cm^{-3}}$, which is typical for a cool-core cluster. The core entropy is $17^{+2}_{-9}$ $\rm{keV~cm^2}$, which is within the lowest ten percent for cluster cores from the ACCEPT samples~\citep{Cavagnolo2009}. The low entropy core is a clear signature of strong cooling in the center of the cluster.
	
	\item The X-ray morphology of CHIPS1911+4455 is highly peaked in the center (96th percentile for ``peakiness'') and very asymmetric (7th percentile for ``symmetry'') compared to a large sample of X-ray-selected clusters. This contradiction in its morphology between a relaxed (peaky) cluster and a dynamically active (asymmetric) cluster is highly unusual, and is rarely observed.
	
	\item Based on the [O\,\textsc{ii}] emission line luminosity, we measure a SFR in the BCG of $189^{+25}_{-22}\,\rm{M_{\odot}\,yr^{-1}}$. This is consistent with an estimate based on the mid-IR continuum of $143^{+31}_{-26}\,\rm{M_{\odot}\,yr^{-1}}$. This BCG is among the five most star-forming BCGs in the low-z Universe. Data from \emph{Hubble} confirm that this emission is extended in complex blue filaments near the BCG, with no evidence for an ongoing merger.
	
	\item Based on the highly asymmetric X-ray morphology on small ($\sim$20\,kpc) scales, coupled with the exceptionally low core entropy, we propose that rapid cooling in this system may have been triggered by a dynamical interaction between two similar-mass clusters. In this scenario, some of the low-entropy gas from the more massive cluster is dislodged from the central AGN-hosting galaxy. For the low-entropy gas separated from the central galaxy, cooling is more favorable due to the longer mixing times, enhanced large-scale turbulence and CCA rain, and lack of a direct heat source (AGN). This system may provide a link to high-redshift clusters, where previous studies have found an abundance of star-forming BCGs in dynamically-active clusters.
\end{enumerate}

CHIPS1911+4455 is the first low-redshift ($z<1$) galaxy cluster with this distinctive characteristic (hosting a high star-forming BCG and a strong cool-core but having a disturbed morphology). The cluster was discovered by the Clusters Hiding in Plain Sight (\emph{CHiPS}) survey because of its exceptionally bright cool core that appears to be point-like in previous X-ray cluster catalogs~\citep{Somboonpanyakul2018}. CHIPS1911+4455 represents a unique opportunity to understand the relationship between a merging galaxy cluster and star formation in its BCG, which, in turn, unravels an alternative method to form cooling flows and massive starbursts apart from a simple accretion model. This mechanism will become much more important at high redshift ($z>1$) when the cluster merger rate is significantly higher~\citep{Fakhouri2010,McDonald2016}. 

\section*{Acknowledgements}
T.\ S.\ and M.\ M.\ acknowledge support from the Kavli Research Investment Fund at MIT, Chandra Award Number GO9-20116X, and by Hubble Award Number HST-GO-16038. M.\ G.\ acknowledges partial support by NASA Chandra GO8-19104X/GO9-20114X and HST GO-15890.020-A. H.\ D.\ and E.\ R.\-T.\ acknowledge support from the Research Council of Norway.

\bibliographystyle{yahapj}

\begin{thebibliography}{}
	\bibitem[B{\^\i}rzan et al.(2008)]{Birzan2008} B{\^\i}rzan, L., McNamara, B.~R., Nulsen, P.~E.~J., et al.\ 2008, \apj, 686, 859
	\bibitem[Bonaventura et al.(2017)]{Bonaventura2017} Bonaventura, N.~R., Webb, T.~M.~A., Muzzin, A., et al.\ 2017, \mnras, 469, 1259
	\bibitem[Burns et al.(1997)]{Burns1997} Burns, J.~O., Loken, C., Gomez, P., et al.\ 1997, Galactic Cluster Cooling Flows, 21
	\bibitem[Burns et al.(2008)]{Burns2008} Burns, J.~O., Hallman, E.~J., Gantner, B., et al.\ 2008, \apj, 675, 1125
	\bibitem[Calzetti et al.(2000)]{Calzetti2000} Calzetti, D., Armus, L., Bohlin, R.~C., et al.\ 2000, ApJ, 533, 682
	\bibitem[Cavagnolo et al.(2008)]{Cavagnolo2008} Cavagnolo, K.~W., Donahue, M., Voit, G.~M., et al.\ 2008, \apjl, 683, L107
	\bibitem[Cavagnolo et al.(2009)]{Cavagnolo2009} Cavagnolo, K.~W., Donahue, M., Voit, G.~M., et al.\ 2009, \apjs, 182, 12
	\bibitem[Cerulo et al.(2019)]{Cerulo2019} Cerulo, P., Orellana, G.~A., \& Covone, G.\ 2019, \mnras, 487, 3759
	\bibitem[Cluver et al.(2017)]{Cluver2017} Cluver, M.~E., Jarrett, T.~H., Dale, D.~A., et al.\ 2017, \apj, 850, 68
	\bibitem[Crawford et al.(1999)]{Crawford1999} Crawford, C.~S., Allen, S.~W., Ebeling, H., et al.\ 1999, MNRAS, 306, 857
	\bibitem[Donahue et al.(2010)]{Donahue2010} Donahue, M., Bruch, S., Wang, E., et al.\ 2010, ApJ, 715, 881
	\bibitem[Edge et al.(1994)]{Edge1994} Edge, A.~C., Fabian, A.~C., Allen, S.~W., et al.\ 1994, \mnras, 270, L1
	\bibitem[Fabian(1994)]{Fabian1994} Fabian, A.~C.\ 1994, \araa, 32, 277
	\bibitem[Fabian(2012)]{Fabian2012} Fabian, A.~C.\ 2012, \araa, 50, 455
	\bibitem[Fakhouri et al.(2010)]{Fakhouri2010} Fakhouri, O., Ma, C.-P., \& Boylan-Kolchin, M.\ 2010, MNRAS, 406, 2267
	\bibitem[Gaspari et al.(2019)]{Gaspari2019} Gaspari, M., Eckert, D., Ettori, S., et al.\ 2019, \apj, 884, 169.
	\bibitem[Gaspari et al.(2020)]{Gaspari2020} Gaspari, M., Tombesi, F., \& Cappi, M.\ 2020, Nature Astronomy, 4, 10.
	\bibitem[G{\'o}mez et al.(2002)]{Gomez2002} G{\'o}mez, P.~L., Loken, C., Roettiger, K., et al.\ 2002, ApJ, 569, 122
	\bibitem[Hlavacek-Larrondo et al.(2015)]{Hlavacek-Larrondo2015} Hlavacek-Larrondo, J., McDonald, M., Benson, B.~A., et al.\ 2015, \apj, 805, 35
	\bibitem[Hudson et al.(2010)]{Hudson2010} Hudson, D.~S., Mittal, R., Reiprich, T.~H., et al.\ 2010, A\&A, 513, A37
	\bibitem[Kennicutt(1998)]{Kennicutt1998} Kennicutt, R.~C.\ 1998, \araa, 36, 189
	\bibitem[Lau et al.(2017)]{Lau2017} Lau, E.~T., Gaspari, M., Nagai, D., et al.\ 2017, \apj, 849, 54.
	\bibitem[Mantz et al.(2015)]{Mantz2015a} Mantz, A.~B., Allen, S.~W., Morris, R.~G., et al.\ 2015, MNRAS, 449, 199
	\bibitem[McDonald et al.(2011)]{McDonald2011} McDonald, M., Veilleux, S., \& Mushotzky, R.\ 2011, ApJ, 731, 33
	\bibitem[McDonald et al.(2012)]{McDonald2012} McDonald, M., Bayliss, M., Benson, B.~A., et al.\ 2012, Nature, 488, 349	
	\bibitem[McDonald et al.(2013)]{McDonald2013} McDonald, M., Benson, B.~A., Vikhlinin, A., et al.\ 2013, ApJ, 774, 23
	\bibitem[McDonald et al.(2016)]{McDonald2016} McDonald, M., Stalder, B., Bayliss, M., et al.\ 2016, ApJ, 817, 86
	\bibitem[McDonald et al.(2017)]{McDonald2017} McDonald, M., Allen, S., Bayliss, M., et al.\ 2017, ApJ, 843, 28
	\bibitem[McDonald et al.(2018)]{McDonald2018} McDonald, M., Gaspari, M., McNamara, B.~R., et al.\ 2018, \apj, 858, 45
	\bibitem[McDonald et al.(2019)]{McDonald2019} McDonald, M., McNamara, B.~R., Voit, G.~M., et al.\ 2019, ApJ, 885, 63
	\bibitem[McNamara et al.(2006)]{McNamara2006} McNamara, B.~R., Rafferty, D.~A., B{\^\i}rzan, L., et al.\ 2006, \apj, 648, 164
	\bibitem[McNamara \& Nulsen(2012)]{McNamara2012} McNamara, B.~R. \& Nulsen, P.~E.~J.\ 2012, New Journal of Physics, 14, 055023
	\bibitem[Menanteau et al.(2012)]{Menanteau2012} Menanteau, F., Hughes, J.~P., Sif{\'o}n, C., et al.\ 2012, ApJ, 748, 7
	\bibitem[Molendi et al.(2016)]{Molendi2016} Molendi, S., Tozzi, P., Gaspari, M., et al.\ 2016, \aap, 595, A123.
	\bibitem[Olivares et al.(2019)]{Olivares2019} Olivares, V., Salome, P., Combes, F., et al.\ 2019, \aap, 631, A22. doi:10.1051/0004-6361/201935350
	\bibitem[Panagoulia et al.(2014)]{Panagoulia2014} Panagoulia, E.~K., Fabian, A.~C., \& Sanders, J.~S.\ 2014, MNRAS, 438, 2341
	\bibitem[Poole et al.(2008)]{Poole2008} Poole, G.~B., Babul, A., McCarthy, I.~G., et al.\ 2008, MNRAS, 391, 1163
	\bibitem[Rafferty et al.(2008)]{Rafferty2008} Rafferty, D.~A., McNamara, B.~R., \& Nulsen, P.~E.~J.\ 2008, \apj, 687, 899
	\bibitem[Somboonpanyakul et al.(2018)]{Somboonpanyakul2018} Somboonpanyakul, T., McDonald, M., Lin, H.~W., et al.\ 2018, ApJ, 863, 122
	\bibitem[Sun(2009)]{Sun2009} Sun, M.\ 2009, \apj, 704, 1586
	\bibitem[Sutherland \& Dopita(1993)]{Sutherland1993} Sutherland, R.~S., \& Dopita, M.~A.\ 1993, ApJS, 88, 253
	\bibitem[Tonry et al.(2012)]{Tonry2012} Tonry, J.~L., Stubbs, C.~W., Lykke, K.~R., et al.\ 2012, \apj, 750, 99 
	\bibitem[Tremblay et al.(2015)]{Tremblay2015} Tremblay, G.~R., O'Dea, C.~P., Baum, S.~A., et al.\ 2015, \mnras, 451, 3768
	\bibitem[Vikhlinin et al.(2006)]{Vikhlinin2006} Vikhlinin, A., Kravtsov, A., Forman, W., et al.\ 2006, \apj, 640, 691
	\bibitem[Voit et al.(2015)]{Voit2015} Voit, G.~M., Donahue, M., Bryan, G.~L., et al.\ 2015, Nature, 519, 203
	\bibitem[Voit et al.(2017)]{Voit2017} Voit, G.~M., Meece, G., Li, Y., et al.\ 2017, \apj, 845, 80.
	\bibitem[Voit(2018)]{Voit2018} Voit, G.~M.\ 2018, \apj, 868, 102.
	\bibitem[Wright et al.(2010)]{Wright2010} Wright, E.~L., Eisenhardt, P.~R.~M., Mainzer, A.~K., et al.\ 2010, AJ, 140, 1868
\end{thebibliography}

\end{document}